\documentclass[10pt]{article}

\usepackage{graphicx}
\usepackage{latexsym}
\usepackage{amsfonts,amsmath,amssymb}
\usepackage[colorlinks=false]{hyperref}
\usepackage[utf8]{inputenc}
\usepackage{fancyref}

\usepackage{cite}

\hypersetup{colorlinks=true,pdfborder={0 0 0},citecolor=black}

\usepackage[labelfont=bf,labelsep=period,justification=raggedright]{caption}

\date{}

\begin{document}

\begin{flushleft}
{\LARGE
\textbf{10 Simple Rules for the Care and Feeding of Scientific Data}
}
\vspace*{1em}
\\
Alyssa Goodman$^{1}$, Alberto Pepe$^{1,*}$, Alexander W. Blocker$^{4}$, Christine L. Borgman$^{2}$, Kyle Cranmer$^{3}$, Merce Crosas$^{4}$, Rosanne Di Stefano$^{1}$, Yolanda Gil$^{5}$, Paul Groth$^{6}$, Margaret Hedstrom$^{7}$, David W. Hogg$^{3}$, Vinay Kashyap$^{1}$,  Ashish Mahabal$^{8}$, Aneta Siemiginowska$^{1}$, Aleksandra Slavkovic$^{9}$ 
\\\vspace*{1em}
{\bf 1} Harvard-Smithsonian Center for Astrophysics, Harvard University
\\
{\bf 2} University of California, Los Angeles (UCLA)
\\
{\bf 3} New York University (NYU)
\\
{\bf 4} Harvard University
\\
{\bf 5} University of Southern California
\\
{\bf 6} VU University Amsterdam
\\
{\bf 7} University of Michigan
\\
{\bf 8} California Institute of Technology (Caltech)
\\
{\bf 9} Penn State University 
\\
{\bf *} Corresponding author: Alberto Pepe (alberto.pepe@gmail.com) 
\\
\end{flushleft}


\section*{Introduction}
In the early 1600s, Galileo Galilei turned a telescope toward Jupiter.
In his log book each night, he drew to-scale schematic diagrams of
Jupiter and some oddly-moving points of light near it. Galileo labeled
each drawing with the date. Eventually he used his observations to
conclude that the Earth orbits the Sun, just as the four Galilean moons
orbit Jupiter. History shows Galileo to be much more than an
astronomical hero, though. His clear and careful record keeping and
publication style not only let Galileo understand the Solar System, it
continues to let \emph{anyone} understand \emph{how} Galileo did it.
Galileo's notes directly integrated his \textbf{data} (drawings of
Jupiter and its moons), key \textbf{metadata} (timing of each
observation, weather, telescope properties), and \textbf{text}
(descriptions of methods, analysis, and conclusions). Critically, when
Galileo included the information from those notes in \emph{Siderius
Nuncius} \cite{galilei}, this integration of text, data and metadata was
preserved, as shown in Figure 1. Galileo's work advanced the
``Scientific Revolution,'' and his approach to observation and analysis
contributed significantly to the shaping of today's modern ``Scientific
Method'' \cite{galilei1618assayer,galilei1957discoveries}.

Today most research projects are considered complete when a journal
article based on the analysis has been written and published. Trouble
is, unlike Galileo's report in \emph{Siderius Nuncius}, the amount of
real data and data description in modern publications is almost never
sufficient to repeat or even statistically verify a study being
presented. Worse, researchers wishing to build upon and extend work
presented in the literature often have trouble recovering data
associated with an article after it has been published. More often than
scientists would like to admit, they cannot even recover the data
associated with their own published works.

Complicating the modern situation, the words ``data'' and ``analysis''
have a wider variety of definitions today than at the time of Galileo.
Theoretical investigations can create large ``data'' sets through
simulations (e.g.
\href{http://www.mpa-garching.mpg.de/galform/virgo/millennium/}{The
Millennium Simulation Project}). Large scale data collection often takes
place as a community-wide effort (e.g.
\href{http://www.genome.gov/10001772}{The Human Genome project}), which
leads to gigantic online ``databases'' (organized collections of data).
Computers are so essential in simulations, and in the processing of
experimental and observational data, that it is also often hard to draw
a dividing line between ``data'' and ``analysis'' (or ``code'') when
discussing the care and feeding of ``data.'' Sometimes, a copy of the
code used to create or process data is so essential to the use of those
data that the code should almost be thought of as part of the
``metadata'' description of the data. Other times, the code used in a
scientific study is more separable from the data, but even then, many
preservation and sharing principles apply to code just as well as they
do to data.

So how do we go about caring for and feeding data? Extra work, no doubt,
is associated with nurturing your data, but care up front will save time
and increase insight later. Even though a growing number of researchers,
especially in large collaborations, know that conducting research with
sharing and reuse in mind is essential, it still requires a paradigm
shift. Most people are still motivated by piling up publications and by
getting to the next one as soon as possible. But, the more we scientists
find ourselves wishing we had access to extant but now unfindable data
\cite{holdren}, the more we will realize why bad data management is bad
for science. How can we improve?

\textbf{This article offers a short guide to the steps scientists can
take to ensure that their data and associated analyses continue to be of
value and to be recognized.} In just the past few years, hundreds of
scholarly papers and reports have been written on questions of data
sharing, data provenance, research reproducibility, licensing,
attribution, privacy, and more--but our goal here is \emph{not} to
review that literature. Instead, we present a short guide intended for
researchers who want to know why it is important to ``care for and
feed'' data, with some practical advice on how to do that. The final section
 at the close of this work offer links to the types of
services referred to throughout the text. \textbf{Boldface lettering}
below highlights actions one can take to follow the suggested rules.

\begin{figure}[h!]
\begin{center}
\includegraphics[width=1\columnwidth]{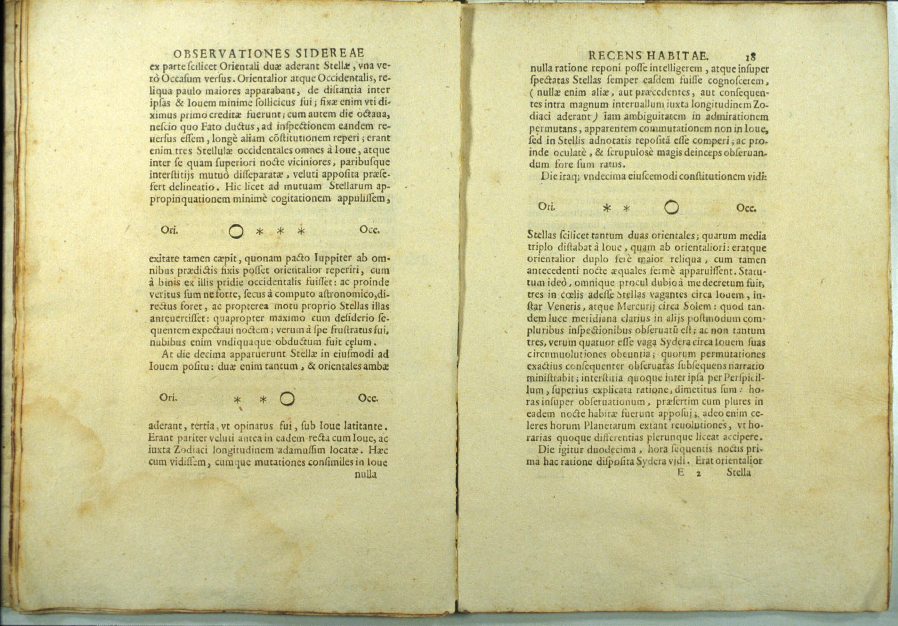}
\caption{Two pages (scan) from Galilei's Sidereus Nuncius (``The Starry Messenger" or ``The Herald of the Stars"), Venice, 
1610. On these pages, Galilei combines data (drawings of Jupiter and its moons), key metadata (timing of each observation, weather, telescope properties), and text (descriptions of methods, analysis, and conclusions). }
\end{center}
\end{figure}

\section*{Rule 1. Love your data, and help others love it too.}

Data management is a repeat-play game. If you take care to make your
data easily available to others, others are more likely to do the
same--eventually. While we wait for this new sharing-equilibrium to be
reached, you can take two important actions. First, cherish, document,
and \textbf{publish your data}, preferably using the robust methods
described in Rule 2. Get started now, as: better tools and resources for
data management are becoming more numerous; universities and research
communities are moving toward bigger investments in data repositories
(Rule 8); and more librarians and scientists are learning data
management skills (Rule 10). At the very least, loving your own data
available will serve \emph{you}: you'll be able to find and reuse your
own data if you treat them well. Second, enable and \textbf{encourage
others to cherish, document, and publish their data}. If you are a
research scientist, chances are that not only are you an author, but
also a reviewer for a specialized journal or conference venue. As a
reviewer, \textbf{request that the authors of papers you review provide
documentation and access to their data} according to the rules set out
in the remainder of this article. While institutional approaches are
clearly essential (Rules 8 and 10), changing minds one scientist at a
time is effective as well.

\section*{Rule 2. Share your data online, with a permanent identifier.}

Nothing really lasts forever, so ``permanent'' actually just means
long-lasting. For example, your personal web site is unlikely to be a
good option for long-term data storage (yet, in the very short run,
putting your data on your site is better than doing nothing at all!). In
general, although many papers include URLs to give access to datasets,
most become inaccessible within a few years \cite{Wren_2008}. The best
option for releasing your data with long-term guarantee is to
\textbf{deposit them in whatever data archive is the ``go to'' place for
your field}. A proper, trustworthy, archive will: (1) assign an
identifier such as a ``handle'' (hdl) or ``digital object identifier''
(doi); (2) require that you provide adequate documentation and metadata;
and (3) manage the ``care and feeding'' of your data by employing good
curation practices. If no such archive exists in your field, there are
also generic (non-domain-specific) online services that can host your
data and issue persistent identifiers (see Rule 8). Pointers to a few
generic repositories are listed in the Resources section (A), and longer compilations
of such services are at the links in Resources (B).

\section*{Rule 3. Conduct science with a particular level of reuse in
mind.}

Data from others are hard to use without context describing what the
data are and how they were obtained. The
\href{http://www.w3.org/TR/2013/REC-prov-dm-20130430/\#dfn-provenance}{W3C
Provenance Group} defines information \emph{provenance} as the sum of
all of the processes, people (institutions or agents), and documents
(data included!) that were involved in generating or otherwise
influencing or delivering a piece of information. Perfect documentation
of provenance is rarely, if ever, attained in scientific work today. The
higher the quality of provenance information, the higher the chance of
enabling data reuse. In general, data reuse is most possible when: 1)
data; 2) metadata (information describing the data); and 3) information
about the process of generating those data, such as code, are all
provided. In trying to follow the Rules listed in this article, you will
do best if you plan in advance for ways to provide all three kinds of
information. \textbf{In carrying out your work, consider what level of
reuse you realistically expect and plan accordingly.} Do you want your
work to be fully \emph{reproducible}? If so, then provenance information
is a must (e.g., working pipeline analysis code, a platform to run it
on, and verifiable versions of the data). Or do you just want your work
to be \emph{inspectable}? If so, then intermediate data products and
pseudo-code may be sufficient. Or maybe your goal is that your data is
\emph{usable} in a wide range of applications? If so, \textbf{consider
adopting standard formats and metadata standards early on}. At the very
least, \textbf{keep careful track of versions of data and code}, with
associated dates. Taking these steps as you plan and carry out projects
will earn you the thanks of researchers, including you, looking back
from the future. (Consult the links in Resources (E) for a list of tools to package all
your research materials with reuse in mind)

\section*{Rule 4. Publish workflow as context.}

Publishing a description of your processing steps offers essential
context for interpreting and re-using data. As-such, scientists
typically include a ``methods'' and/or ``analysis'' section(s) in a
scholarly article, used to describe data collection, manipulation, and
analysis processes. Computer and information scientists call the
combination of the collection methods and analysis processes for a
project its ``workflow,'' and they consider the information used and
captured in workflow to be part of the ``provenance'' of the data. In
some cases (mostly in genomics), scientists can use existing workflow
software in \emph{running} experiments and in \emph{recording} what was
done in those experiments, e.g. \href{www.genepattern.org}{Gene
Pattern}. In that best-case scenario, the workflow software, its
version, and settings used can be published alongside data using the
other rules laid out here. But, it is rare outside of genomics to see
the end-to-end process described in a research paper run, orchestrated,
and/or recorded by a single software package. In a plausible utopian
future, automated workflow documentation could extend to all fields, so
that an electronic provenance record could link together all the pieces
that led to a result: the data citation (Rule 2), the pointer to the
code (Rule 6), the workflow (this Rule), and a scholarly paper (Rule 5).
But what can you do now? \textbf{At a minimum, provide, alongside any
deposit of data, a simple sketch of data flow across software,
indicating how intermediate and final data products and results are
generated. If it's feasible and you are willing to deal with a higher
level of complexity, also consider using an online service to
encapsulate your workflow (see Resources (C) for a list of services)}.
Keep in mind that even if the data used are not ``new,'' in that they
come from a well-documented archive, it is still important to document
the archive query that produced the data you used, along with all the
operations you performed on the data after they were retrieved. Keeping
better track of workflow, as context, will likely benefit you and your
collaborators enough to justify the loftier, more altruistic, goals
espoused here.

\section*{Rule 5. Link your data to your publications as often as
possible.}

Whether your ``data'' include tables, spreadsheets, images, graphs,
databases and/or code, you should make as much of it as possible
available \emph{with} any paper that presents it. \textbf{If it's
practical and helpful, share your data as early as possible in your
research workflow: as soon as you are done with the analysis, even
before you write any article(s) about it.} Your data can even be cited
before (or without) its inclusion in a paper (see Rule 7). Many journals
now offer standard ways to contribute data to their archives and link it
to your paper, often with a persistent identifier. Whenever possible,
\textbf{embed citations (links) to your data and code, each with its own
persistent identifier, right into the text of your paper, just like you
would reference other literature.} If a journal hosting your paper
doesn't offer a place for your data, and or an identifier for it, use a
repository (Rule 8) and get your own identifier (Rule 2). At a
\emph{minimum}, you can post, and refer to, a package of files (data,
codes, documentation on parameters, metadata, license information,
and/or lists of links to such) with a persistent online identifier (Rule
2). And, if your domain's journals' policies do not allow for good
data-literature interlinking, try to effect change (see Rules 1 and 10).

\section*{Rule 6. Publish your code (even the small bits).}

Did you write any code to run your analysis? \textbf{No matter how buggy
and insignificant you may find it, publish it.} Many easy-to-use source
code repositories exist, which allow not only hosting of software but
also facilitate collaboration and version tracking (see Resources, D).
Your code, even the shortest script (whether or not you are proud of its
quality), can be an important component for understanding your data and
how you got your results \cite{Barnes_2010}. Software plays several
roles in relation to data and scientific research, and norms around its
publication are still evolving and different across disciplines
\cite{iroff_Mink_Hanisch_DuPrie_2013}. In some cases, software is the
primary data product (e.g., new algorithms). In some cases, software is
the primary data product (e.g., new algorithms). In some other cases,
data are the primary research products, yet the best way to document
their provenance is to publish the software that was used to generate
them as ``metadata.'' In both cases, publishing the source code and its
version history is crucial to enhance transparency and reproducibility.
The use of open source software when possible reduces barriers for
subsequent users of your software related data products.
\cite{Prlic_Procter_2012} The same best practices discussed above in
relation to data and workflow also apply to software materials: cite the
software that you use and provide unique, persistent identifiers (Rule
2) to the code you share.

\section*{Rule 7. Say how you want to get credit.}

Chances are that you want to get credit for what you share. The
attribution system used for scholarly articles, accomplished via
citations, often breaks in the case of data and software. When other
authors reuse or cite your data or code, you may get an acknowledgement
or an incoming link. If you and your colleagues have gone to the trouble
to write a ``data paper,'' whose main purpose is to describe your data
and/or code, you may also get a citation. \cite{Chavan_Penev_2011} But,
``data paper'' writing is not always desirable, or relevant. So, how do
you go about getting the full credit you deserve for your data and code?
\textbf{The best way is to simply describe your expectations on how you
would like to be acknowledged.} If you want, \textbf{you can also
release your data under a license and indicate explicitly in the paper
or in the metadata how you want others to give you credit.} But, while
legal mechanisms have advantages, they can also inadvertently lead to
limitations on the reuse of the data you are sharing. In any case,
\textbf{make information about you (e.g.~name, institution), about the
data and/or code (e.g.~origin, version, associated files and metadata),
and about exactly how you would like to get credit, as clear as
possible.} Easy-to-implement licenses, many of which offer the advantage
of being machine-readable, are offered by the
\href{http://creativecommons.org/choose/}{Creative Commons}
organization, as are other similar options, such as those offered by
\href{http://opendatacommons.org/licenses/pddl/}{Open Data Commons}.
Resources section, G, provides more information.

\section*{Rule 8. Foster and use data repositories.}

Sometimes the hardest and most time-consuming step of sharing data and
code is finding and deciding where to put them. Data-sharing practices
vary widely across disciplines: in some fields data sharing and reuse
are essential and commonplace, while in others data sharing is a ``gift
exchange'' culture \cite{Wallis_Rolando_Borgman_2013}. \textbf{If your
community already has a standard repository, use it. If you don't know
where to start looking, or you need help choosing amongst relevant
repositories, ask an information specialist, such as a data scientist or
a librarian working in your field} (and consult the directories of data
repositories listed in Resurces, B). When choosing amongst repositories,
try to find the one offering the best combination ease-of-deposit,
community uptake, accessibility, discoverability, value-added curation,
preservation infrastructure, organizational persistence, and support for
the data formats and standards you use. \textbf{Remember that even if
your field has no domain-based repository, your institution may have
one}, and your local librarian or archivist can instruct you on how to
use that local resource. If neither your community nor your institution
has a relevant repository, try a generic repository or consider setting
up your own (see Rule 2, and Resources, F).

\section*{Rule 9. Reward colleagues who share their data properly.}

Whether you do it in person at scientific meetings and conferences or by
written communication when reviewing papers and grants, \textbf{reward
your colleagues who share data and code. Rally your colleagues and
engage your community by providing feedback on the quality of the data
assets in your field. Praise those following the best practices.} The
more the data created by your colleagues is accessible as an organized
collection of some sort, the better your community's research capacity.
The more data get shared, used, and cited, the more they improve.
Besides personal involvement and encouragement, the best way to reward
data sharing is by attribution: always cite the sources of data that you
use. \textbf{Follow good scientific practice and give credit to those
whose data you use, following their preferred reference format and
according to current best practices.} Standards and practices for citing
and attributing data sources are actively being developed through
international partnerships. \cite{national2012For,manifesto}

\section*{Rule 10: Be a booster for data science.}

As Rule 1 says, it is important not just that \emph{you} love your own
data, but that \emph{others} love data too. An attitude that data and
code are ``2nd class objects,'' behind traditional scholarly
publications is still prevalent. But, every day, as scientists try to
use the frustrating but tantalizing hodgepodge of research data
available via the present ad-hoc network of online systems, the value of
organizing an open network of re-usable data and code is becoming more
and more clear, to more and more people. \textbf{You, as a scientist,
need to help organize your discipline, and your institution to move more
quickly to a world of open, discoverable, reproducible data and
research. One important step is to \emph{advocate} for hiring data
specialists and for the overall support of institutional programs that
improve data sharing.} Make sure not only advanced researchers (e.g.,
postdocs) experience the pleasures of doing research with freely
available data and tools: \textbf{\emph{explain} and \emph{show} the
value of well-loved data to graduate and undergraduate researchers}.
\emph{Teach} whole courses, or mini-courses, related to caring for data
and software, or incorporate the ideas into existing courses. \emph{Form
groups} specific to your discipline to foster data and code sharing.
Hold birds-of-a-feather or special \emph{sessions during large meetings}
demonstrating examples where good sharing practices have led to better
results and collaborations. Lead by practicing what you preach.

\section*{Links to Useful Resources}

\subsection*{A: General Data Repositories}

\begin{itemize}
\item
  \href{http://thedata.org}{Dataverse} A repository for research data
  that takes care of long term preservation and good archival practices,
  while researchers can share, keep control of and get recognition for
  their data.
\item
  \href{http://figshare.com}{FigShare} A repository where users can make
  all of their research outputs available in a citable, shareable and
  discoverable manner.
\item
  \href{http://zenodo.org}{Zenodo} A repository service that enables
  researchers, scientists, projects and institutions to share and
  showcase multidisciplinary research results (data and publications)
  that are not part of existing institutional or subject-based
  repositories.
\item
  \href{http://dryad.org}{Dryad} A repository that aims to make data
  archiving as simple and as rewarding as possible through a suite of
  services not necessarily provided by publishers or institutional
  websites.
\end{itemize}
\subsection*{B: Directories of Research Data Repositories}

\begin{itemize}
\item
  \href{http://databib.org}{DataBib}. Databib is a tool for helping
  people identify and locate online repositories of research data. Users
  and bibliographers create and curate records that describe data
  repositories that users can search.
\item
  \href{http://http://www.re3data.org/}{re3data.org}. Re3data is a
  global registry of research data repositories from different academic
  disciplines for researchers, funding bodies, publishers and scholarly
  institutions.
\item
  \href{http://oad.simmons.edu/oadwiki/Data\_repositories}{Data
  repositories, Open Access Directory} A list of repositories and
  databases for open data.
\item
  \href{http://www.force11.org/catalog}{Force 11 Catalog} A dynamic
  inventory of web-based scholarly resources, a collection of
  alternative publication systems, databases, organizations and groups,
  software, services, standards, formats, and training tools.
\end{itemize}

\subsection*{C: Workflow Management Systems}

\begin{itemize}
\item
  \href{http://www.taverna.org.uk/}{Taverna} An open source and
  domain-independent Workflow Management System -- a suite of tools used
  to design and execute scientific workflows and aid in silico
  experimentation.
\item
  \href{https://kepler-project.org/}{Kepler} Software designed to help
  scien{\-}tists, analysts, and computer programmers create, execute, and
  share models and analyses across a broad range of scientific and
  engineering disciplines.
\item
  \href{http://www.wings-workflows.org/}{Wings} A semantic workflow
  system that assists scientists with the design of computational
  experiments.
\item
  \href{http://www.vistrails.org/index.php/Main\_Page}{VisTrails} An
  open-source scientific workflow and provenance management system that
  supports data exploration and visualization.
\item
  \href{http://www.knime.org/}{Knime} A graphical workbench for the
  entire analysis process: data access, data transformation, initial
  investigation, powerful predictive analytics, visualization and
  reporting.
\end{itemize}

\subsection*{D: Source Code Repositories}

\begin{itemize}
\item
  \href{http://github.com}{Github} A web-based hosting service for
  software development projects that use the Git revision control
  system, including many open source projects.
\item
  \href{http://git-scm.com/}{Git} A free and open source distributed
  version control system designed to handle everything from small to
  very large projects with speed and efficiency.
\item
  \href{http://mercurial.selenic.com/}{Mercurial} A free, distributed
  source control management tool. It efficiently handles projects of any
  size and offers an easy and intuitive interface.
\item
  \href{https://bitbucket.org/}{BitBucket} A web-based hosting service
  for projects that use either the Mercurial or Git revision control
  systems.
\end{itemize}

\subsection*{E: Systems to Package, Access, and Execute Data and
Code}

\begin{itemize}
\item
  \href{http://ipython.org/notebook.html}{iPython Notebooks} A web-based
  interactive computational environment where you can combine code
  execution, text, mathematics, plots and rich media into a single
  document.
\item
  \href{http://ropensci.org/}{ROpenSci} A suite of packages that allow
  access to data repositories through the R statistical programming
  environment.
\item
  \href{https://authorea.com}{Authorea} A collaborative online word
  processor for scholarly papers that allows the writing of web-native,
  living, dynamic, ``executable'' articles which include text,
  mathematical notation, images, and data. It currently support
  inclusion and rendering of d3.js and iPython notebooks.
\item
  \href{http://dexy.it}{Dexy} A multi-purpose project automation tool
  for working with documents via a command-line interface.
\end{itemize}

\subsection*{F: Software Tools to Run Your Own Document
Repository}

\begin{itemize}
\item
  \href{http://invenio-software.org/}{Invenio}. Invenio is a {}free
  software suite enabling you to run your own {}digital library or
  document repository on the web. Invenio is an ideal solution to run
  document repositories of moderate to large sizes (several millions of
  records). Invenio is co-developed by CERN, {}DESY, {}EPFL, {}FNAL, {}and
  SLAC.
\item
  \href{http://www.eprints.org/software/}{Eprints}. EPrints is one of
  the easiest and fastest way to set up small to medium sized
  repositories of open access research literature, scientific data,
  theses, reports and multimedia. Developed at the University of
  Southampton, UK.
\item
  \href{http://www.dspace.org/}{DSpace}. DSpace is a turnkey
  institutional repository application developed by the Duraspace
  organization.
\end{itemize}

\subsection*{G: Licensing and Privacy}

\begin{itemize}
\item
  \href{http://opensource.org/licenses}{Open Source Initiative,
  Licenses} Open source licenses are licenses that comply with the Open
  Source Definition: they allow software to be freely used, modified,
  and shared. These include Apache, BSD, GNU (GPL), MIT, and Mozilla
  Public License.
\item
  \href{http://privacytools.seas.harvard.edu}{Privacy Tools for Sharing
  Research Data} A Harvard-based collaborative and multidisciplinary
  effort to help enable the collection, analysis, and sharing of
  personal data for research in social science and other fields while
  providing privacy for individual subjects.
\end{itemize}

\section*{Acknowledgment}

This article was written \textbf{collaboratively}, \textbf{online}, in
the \textbf{open}, using Authorea. All the authors participated in discussions at the the exploratory
seminar which led to the preparation of this article. A. Goodman and
A.Pepe wrote the bulk of the article. C. Borgman, M. Crosas, K. Cranmer,
R. Di Stefano, P. Groth, Y. Gill, M. Hedstrom, A. Mahabal, and A.
Slavkovic contributed to the article with substantial edits. All the
authors provided comments on the various stages of the article.
Every iteration of the writing
procedure is logged and available in the online version of this article
at \href{https://www.authorea.com/3410}{authorea.com/3410}. 

This article
is an outcome of an Exploratory Seminar called \emph{What to Keep and
How to Analyze It: Data Curation and Data Analysis with Multiple Phases}
(\href{http://www.radcliffe.harvard.edu/exploratory-seminars/what-to-keep-and-how-to-analyze-it}{link})
organized by Xiao-Li Meng and Alyssa Goodman, held on May 9-10, 2013 at
the \textbf{Radcliffe Institute for Advanced Study, Harvard University},
Cambridge, Mass.

\bibliography{bibliography/converted_to_latex.bib}

\end{document}